\documentclass[conference]{IEEEtran}

\ifCLASSINFOpdf
\else
\fi

\usepackage{graphicx}
\usepackage{amsmath}
\usepackage{caption}
\usepackage{amsfonts}
\usepackage{amsthm}
\usepackage{algorithm}
\usepackage{algpseudocode}
\usepackage{booktabs} 
\usepackage{subfigure}
\usepackage{color}
\usepackage{epstopdf}
\usepackage[numbers,sort&compress]{natbib}

\begin{document}

\title{\huge Studying the brain from adolescence to adulthood through sparse multi-view matrix factorisations}


\author{
\IEEEauthorblockN{Zi Wang\,$^{1}$, Vyacheslav Karolis\,$^{2}$, Chiara Nosarti\,$^{2,}$\,$^{3}$, and Giovanni Montana\,$^{1}$}
\IEEEauthorblockA{$^{1}$Department of Biomedical Engineering, King's College London, London, SE1 7EH, UK   \\
$^{2}$Department of Psychosis Studies, $^{3}$Centre for the Developing Brain, King's College London, London, SE5 8AF, UK }}

\maketitle

\begin{abstract}

Men and women differ in specific cognitive abilities and in the expression of several neuropsychiatric conditions. Such findings could be attributed to sex hormones, brain differences, as well as a number of environmental variables. Existing research on identifying sex-related differences in brain structure have predominantly used cross-sectional studies to investigate, for instance, differences in average gray matter volumes (GMVs). In this article we explore the potential of a recently proposed multi-view matrix factorisation (MVMF) methodology to study structural brain changes in men and women that occur from adolescence to adulthood. MVMF is a multivariate variance decomposition technique that extends principal component analysis to "multi-view" datasets, i.e. where multiple and related groups of observations are available. In this application, each view represents a different age group. MVMF identifies latent factors explaining shared and age-specific contributions to the observed overall variability in GMVs over time. These latent factors can be used to produce low-dimensional visualisations of the data that emphasise age-specific effects once the shared effects have been accounted for. The analysis of two datasets consisting of individuals born prematurely as well as healthy controls provides evidence to suggest that the separation between males and females becomes increasingly larger as the brain transitions from adolescence to adulthood. We report on specific brain regions associated to these variance effects.

\end{abstract}

\IEEEpeerreviewmaketitle

\bigskip

\section{Introduction}

Men and women differ in specific cognitive abilities and in the expression of several neuropsychiatric conditions. For instance, a male advantage has been suggested for spatial tasks, such as mentally rotating 3D objects \cite{voyer95}, whereas a female advantage has been described in reading \cite{reilly12}. Furthermore, there are considerable differences between men and women in terms of prevalence, age of onset, and symptomatology of several neuropsychiatric disorders \cite{rutter03}. These findings could be attributed to sex hormones, brain differences, and a number of environmental variables, such as cultural values and stereotypes \cite{miller14}. In this paper we try to elucidate the influence of sex on the brain in order to better understand the developmental mechanisms which are associated with sex-biased cognitive abilities and neuropsychiatric conditions. 

A growing body of research uses magnetic resonance imaging (MRI) to explore the role of different brain regions in cognition and behavior. The investigation of longitudinal changes in brain development is particularly important in the light of recent studies, suggesting that dynamic sequences of cortical and subcortical maturation across prolonged time periods, rather than cross-sectional measurements at defined time points, may better discriminate between groups, e.g. males and females \cite{lenroot07}. Results of a meta-analysis of sex differences in human brain structure showed that males had greater voxel-based mean estimates of volume and density compared to females in limbic areas including the amygdala and the hippocampus, and in posterior cingulate gyrus; whereas females had greater volumes/densities in brain regions in the right hemisphere typically associated with language, as well as limbic regions including the insula \cite{ruigrok14}. Other independent investigations showed that variance in regional brain volume was different in separate age groups and typically increased with age \cite{dickie13, kruggel06}.

In this article we investigate whether the difference in neurodevelopmental process between males and females at various stages from adolescence to adulthood could explain the gray matter volume (GMV) variability specific to each age group and shared across all age groups. A simple approach using standard multivariate methodology might consist of performing a principal component analysis (PCA) independently for each age group. This analysis can provide a set of group-specific orthogonal projections explaining a large portion of GMV sample variance, possibly revealing discrimination between males and females. However, carrying out a  multivariate analysis independently on each age group would fail to identify any potential sources of sample variability in observed GMV values that are {\it shared} across all age groups; when such shared factors exist, they should be estimated first, and then discounted, in order to better characterise the data distribution within each age group. Furthermore, treating each age group as independent would fail to capture the correlation patterns that are typically present in longitudinal measurements. 

Motivated by these requirements, we have explored the benefits of a recently proposed multivariate analysis technique for the joint analysis of multiple groups, or "views", called {\it multi-view matrix factorisation} (MVMF) \cite{zw15}. MVMF will allow us to decompose the total GMV sample variance in each age group into the sum of two orthogonal components: one capturing the variance that is shared across all age groups, and one that isolates the age-group specific variances. Furthermore, a regularised version of MVMF can identify specific brain regions contributing to both shared and age-specific variance components. As in PCA, MVMF derives orthogonal coordinate systems to facilitate low-dimensional visualisations of the data. Unlike PCA, these visualisations can be obtained separately for age-shared and age-specific latent factors, thus providing more insights into neurodevelopmental mechanisms. By looking for separation in the low-dimensional cloud points we will be able to assess whether gender-related differences explain a proportion of overall GMV variability, and how this proportion varies with age, from adolescence to adulthood. 

\bigskip

\section{Method}

The methodology presented here assumes a longitudinal dataset consisting of $M$ age groups. In the $m^{th}$ group, we have $n_m$ subjects of similar age who may or may not have repeated measurements in other groups. For each subject we have observed GMVs across $p$ brain regions. The resulting $M$ data matrices are denoted by $X^{(1)}$,..., $X^{(M)}$, where $X^{(m)} \in \mathbb{R}^{n_m \times p}$ so that the rows correspond to subjects and the columns to brain regions. For each $X^{(m)}$, we subtract the column means. The $j^{th}$ diagonal entry of the $p \times p$ gram matrix, $\frac{1}{n_m}(X^{(m)})^TX^{(m)}$, is the sample variance of the $j^{th}$ brain region, and the trace is the total sample variance. 

Our ultimate aim is to identify brain regions that jointly explain a large amount of sample variances across all age groups as well as brain regions that explain age-specific contributions of the overall variance across time. Following \cite{zw15}, our strategy involves approximating each data matrix by the sum of shared and an age-specific components, i.e.:
\begin{equation}  \label{prototype}
\frac{1}{\sqrt{n_m}}X^{(m)} ~  \approx \underbrace{S^{(m)}}_\text{\scriptsize shared component} + ~~~ \underbrace{T^{(m)}}_\text{\scriptsize age-specific component} 
\end{equation}
for $m=1,2,...,M$, where $1/\sqrt{n_m}$ is a scaling factor such that the trace of the gram matrix of the left-hand-side equals the sample variance. These components are required to satisfy three properties: (a) the rank of $S^{(m)}$ and $T^{(m)}$ are both much smaller than min$(n_m,p)$ so that they reveal intrinsic structure of the data while discarding redundant information; (b) the total variance explained by $S^{(m)}$ and $T^{(m)}$ equals the sum of the variance explained by each individual component; (c) the shared component explains the same amount of GMV variance in each brain region in all age groups so that differential GMV variation between age groups is exclusively captured in the age-specific component. 

The methodology presented in \cite{zw15} consists of factorising $S^{(m)}$ and $T^{(m)}$ such that the above properties are satisfied. First, it is assumed that rank$(S^{(m)})=d$ and rank$(T^{(m)})=r$, where $d,r << \text{min}(n_m,p)$ so that dimensionality is greatly reduced. For a given $r$, $T^{(m)}$ can be expressed as:

$$
T^{(m)}=W^{(m)}(V^{(m)})^T=\sum_{j=1}^{r}W^{(m)}_j (V^{(m)}_j)^T=\sum_{j=1}^{r}T^{(m)}_{[j]}
$$
where the subscript $j$ denotes the $j^{th}$ column of the matrix. Each $T^{(m)}_{[j]}$ has the same dimensionality as $T^{(m)}$ and is composed of an age-specific latent factor (LF). For each $j$, $W^{(m)}_j$ is an $n_m$-dimensional vector in a latent space which can then be mapped to the $\mathbb{R}^p$ space using a linear transformation, whose coefficients are defined by the $j^{th}$ row of $V^{(m)}$. An orthogonal constraint, $(W^{(m)})^TW^{(m)}=I_r$, is introduced so that the matrix factorisation is unique subject to an isometric transformation. Analogously, $S^{(m)}$ is factorised as:
\begin{equation}    \label{Sfactorisation}
S^{(m)}=U^{(m)}(V^*)^T=\sum_{k=1}^{d}U^{(m)}_k (V^*_k)^T=\sum_{k=1}^{d}S^{(m)}_{[k]}  
\end{equation} 
where $U^{(m)}$ is orthogonal. The amount of shared variance explained for the $j^{th}$ brain region in age group $m$ is:
\begin{align}
\text{Tr}\{(S^{(m)})^TS^{(m)}\}_j 
&= \text{Tr}\{V^*(V^*)^T\}_j  \nonumber
\end{align}
which is independent of $m$. Each $S^{(m)}_{[k]}$ in \eqref{Sfactorisation} has the same dimensionality as $S^{(m)}$ and is composed of one shared variability LF. In order to ensure that $S^{(m)}$ and $T^{(m)}$ are uncorrelated, a further constrain $(U^{(m)})^TW^{(m)}=0_{d \times r}$ is imposed. As such, $(S^{(m)})^TT^{(m)}=V^*U^{(m)}W^{(m)}V^{(m)}=0_{p \times p}$ and the amount of variance explained in the $m^{th}$ age group is: $\text{Tr}\{(S^{(m)})^TS^{(m)}\} + \text{Tr}\{(T^{(m)})^TT^{(m)}\}$, which yields the required variance decomposition. 

The reconstruction error can then be written as 
$$
\ell = \sum_{m=1}^M \big\Arrowvert \frac{1}{\sqrt{n_m}}X^{(m)} - U^{(m)}(V^*)^{T}-W^{(m)}(V^{(m)})^{T} \big\Arrowvert_\mathcal{F}^2
$$
where $\Arrowvert . \Arrowvert_\mathcal{F}$ denotes the Frobenius norm. MVMF estimates all the parameters by minimizing
\begin{center}
$
\ell + 2 \cdot M \cdot \Arrowvert V^* \Lambda^* \Arrowvert_1 + 2 \sum_{m=1}^{M} \Arrowvert V^{(m)} \Lambda^{(m)} \Arrowvert_1
$
\end{center}
subject to the constraints that the columns of $W^{(m)}$ and $U^{(m)}$ are orthonormal. For our study we used a penalised version of this objective function so as to force some of the coefficients in $V^{(m)}$ and $V^*$ to be exactly zero, and gain insights into the importance of specific brain regions contributing to the variance decompositions. $\Lambda^*$ and $\Lambda^{(m)}$ are $d \times d$ and $r \times r$ diagonal matrices, respectively, where the $k^{th}$ diagonal entry is a non-negative regularisation parameter such that a larger value induces more zero entries in the corresponding column of $V^{(m)}$ and $V^*$. A solution is found by iteratively minimising with respect to $U^{(m)}$ and $W^{(m)}$, $V^*$, $V^{(m)}$ until convergence. 

Furthermore, MVMF can be used to obtain low-dimensional representations of the data for visualisation purposes. Cartesian coordinate systems can be derived whereby each axis represents either a shared or age-specific latent factor. The coordinates of the $i^{th}$ subject in age group $m$ are the corresponding columns of the $i^{th}$ row of $U^{(m)}$ and $W^{(m)}$. We are typically interested in 2- or 3-dimensional projections for visualisation purposes. Denoting the $n_m \times (d+r)$ matrix $[U^{(m)}|W^{(m)}]$ by $\tilde{U}^{(m)}$ and the $p \times (d+r)$ matrix $[V^*|V^{(m)}]$ by $\tilde{V}^{(m)}$, each data matrix can be approximated as $X^{(m)}\approx \tilde{U}^{(m)}\tilde{V}^{(m)}$. The orthonormal matrix $\tilde{U}^{(m)}$ consists of normalised principal projections (PPJs) whereas $\tilde{V}^{(m)}$ contains the loadings.

\bigskip

\section{Materials and results}

\subsection{Dataset}

\begin{figure*}[!tpb]
 \centerline{\includegraphics[scale=0.20]{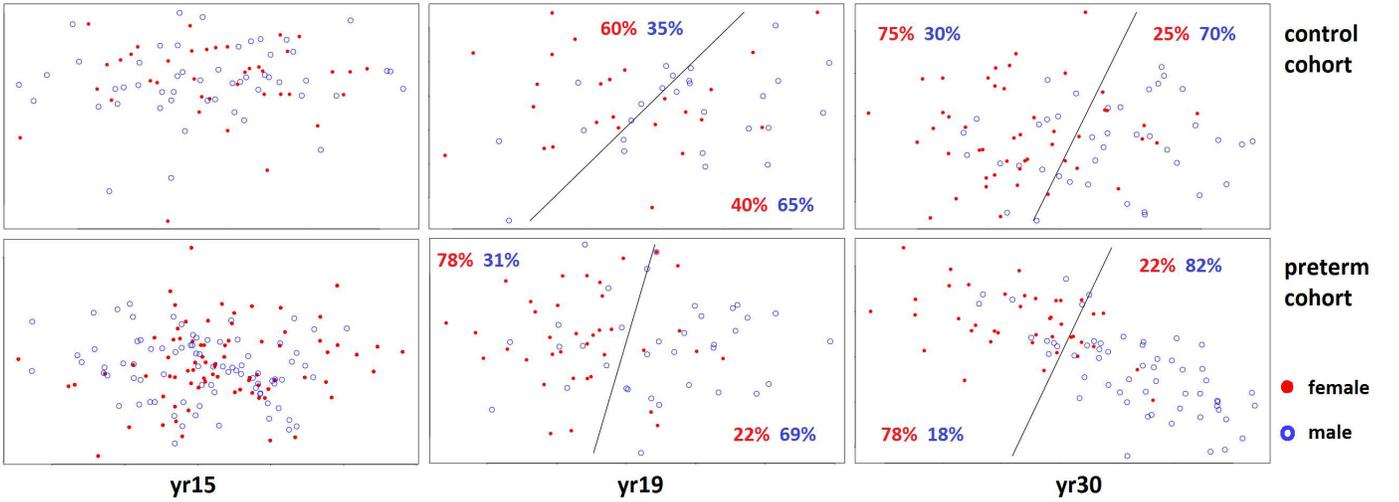}}
 \caption{\small Age-specific PPJs of the control and preterm cohort subjects, colour-coded for sex where red solid circles represent females and blue hollow circles represent males. After removing the mean GMV from each brain region and leaving out the shared GMV variability across all age groups, we see no evident separation between the male and female subjects in the age-specific PPJ plots at the age of $15$, but the separation appears to be present at the age of $19$ and becomes clearer at the age of $30$ in both cohorts. We also plot the linear discriminant analysis (LDA) classification boundary when LDA performs better than random guessing, and report the percentage of each sex falling on each side of the boundary. }
 \label{fig:ppj}

\end{figure*}

MRI data were collected for a preterm cohort and a control group. Participants in the preterm group consisted of individuals born between $1979$ and $1984$ before $33$ weeks of gestation and admitted consecutively to the Neonatal Unit of University College London Hospital. These included $160$ subjects scanned around age $15$ (mean $15.2$, $82$ males), $67$ around age $19$ (mean $20.1$, $30$ males) \cite{nam15}, and $97$ around age $30$ (mean $30.8$, $57$ males). Participants in the control cohort were born full-term ($38$-$42$ weeks) and with a birth weight greater than $2500$ grams. Exclusion criteria included any history of neurological conditions (e.g. meningitis and head injury). In the current analyses $92$ subjects were studied around age $15$ (mean $15.1$, $52$ males), $53$ around age $19$ (mean $19.5$, $26$ males), and $87$ around age $30$ (mean $30.3$, $40$ males). Gray matter probability maps were extracted from T1-weighted images using the unified segmentation approach implemented in the SPM toolbox and registered to a study-specific template created with Advanced Normalization Tools (GreedySyN pipeline). They were subsequently modulated and smoothed in order to produce gray matter volumetric maps. Average GMV per voxel was calculated for $84$ cortical and subcortical regions demarcated by automatized Freesurfer parcellation of the template. The processed GMV of $p=84$ brain regions of $3$ age groups were arranged in $n_m \times p$ matrices $\{X^{(m)}:~m=1,2,3\}$ for the preterm participants and $n'_m \times p$ matrices $\{Y^{(m)}:~m=1,2,3\}$ for the controls. The columns-wise means were centered at zero. 

\subsection{Sex-related differences in variance components}

We fixed $r=2$ and searched for the minimum $d$ such that the model explained no less than $90\%$ of the total variance in at least one age group. Using this procedure, we set $d=20$ in the controls dataset and $d=13$ in the preterm dataset. The shared LFs explained $82\%, 47\%, 70\%$ of total GMV variance at age $15$, $19$, $30$ respectively, in the control cohort. In the preterm cohort, these percentages were $82\%, 39\%, 42\%$. The age-specific LFs explained $9.4\%, 18\%, 14\%$ of total GMV variance at age $15$, $19$, $30$, respectively, in the control cohort. These percentages were $9.7\%, 19\%, 27\%$ in the preterm cohort. In this study we only consider the $r=2$ case as we are interested in visually assessing the projected data in two dimensions in an attempt to explain whether age-specific GMV variances can be driven by differences between males and females.

Age-specific PPJs for both datasets are shown in Figure \ref{fig:ppj}. Looking at the upper panel, we note no evident gender separation in the control cohort derived from the variance contribution at $15$ years of age. However, there appears to be a clearer separation at $19$ years, which then results in a clearly distinguishable linear discrimination at $30$ years. The straight lines are obtained using linear discriminant analysis (LDA) by taking males and females as group labels. We also report on the percentage of subjects within each sex falling on each side of the LDA boundary. Similar patterns can be noted in the lower panel of Figure \ref{fig:ppj} for the preterm cohort. We stress again that these projections are derived after removing the contribution to the overall variance that is not age-specific, and that explains a large amount of variance simultaneously across all age groups. As can be assessed visually, a fair amount of yr$19$-specific variance and a substantial amount of yr$30$-specific variance of GMVs may be related to sexual dimorphism, directing towards the hypothesis that the neurodevelopmental processes of male and female human brains deviate increasingly from adolescence to adulthood. 


\subsection{Age-specific regional differences in variance components}

In order to identify brain regions where sexual dimorphism at the age of $30$ and/or $19$ was most evident, sMVMF was fitted to the preterm and control cohorts separately. We adopted a variable selection procedure, {\it stability selection} \cite{meinshausen10}, to rank the brain regions according to their importance in explaining shared and age-specific variances. This procedure consisted of fitting the sMVMF with fixed sparsity parameters to $10,000$ randomly extracted sub-samples (with replacement), where each sub-sample comprised half of the subjects from each data matrix. For each model fit, we kept track of all the brain regions selection by the model, and eventually used the empirical selection probabilities (SPs) across all sub-samples for all shared and age-specific components. This process was repeated independently for each cohort. 

While the values of SPs are dependent on the specific choice of $\Lambda^{(m)}$ and $\Lambda^*$, the order of the brain regions by their SPs is generally consistent across a wide range of sparsity parameters. In our experiment we set $\Lambda^{(m)}$ and $\Lambda^*$ such that two of the most representative brain regions from each shared and age-specific LF were selected in each sub-sample. Once we obtained the SPs of all brain regions for a cohort, we ranked the brain regions according to their shared or age-specific SPs in descending order. A higher ranking brain region by the shared SP implies that this brain region contains a lot of shared GMV variance across all age groups, and likewise a higher ranking region by yr30-specific SP implies that the region contains a lot variance which is specific amongst the 30-year-olds. We were particularly interested in high-ranking brain regions by yr30-specific and/or yr19-specific SPs in both preterm and control cohorts since they were likely to account for the clustering pattern according to sex at the age of $19$ and $30$ as observed in the middle and right panels in Figure \ref{fig:ppj}. Table \ref{tab:regions} highlights a list of such brain regions in each age group, the ranking, and SPs in both preterm and control cohorts for further investigation.

\begin{table}[!h]  
\caption{\small Top ranking brain regions associated to yr19- and yr30-specific variance components in both datasets. Selection probabilities (SP) are reported in brackets.}    \label{tab:regions}
{\begin{tabular}{llll}
\hline
age & Brain region        & Rank (SP)   & Rank (SP) \\
group    & (L)eft and (R)ight  & in preterms & in controls \\
\hline
19 yrs & supramarginal gyrus (R) & 1 (.852) & 1 (.536) \\
& nucleus accumbens (L) & 1 (.885) & 1 (.887) \\
& temporo-parietal junction (L) & 4 (.413) & 2 (.681) \\
& caudal middle frontal gyrus (L) &  5 (.376) & 6 (.357) \\
& frontal pole (L) & 2 (.721) & 4 (.523) \\
& caudal anterior cingulate gyrus (R) & 3 (.491) & 4 (.281) \\
\hline
30 yrs & temporo-parietal junction (L) & 1 (.901) & 3 (.814) \\
& nucleus accumbens (L) & 2 (.892) & 1 (.960) \\
& amygdala (L) & 5 (.534) & 2 (.956) \\
& hippocampus (R) & 3 (.535) & 1 (.950) \\
& pericalcarine sulcus (L) & 7 (.380) & 4 (.798) \\
& pars triangularis (R) & 5 (.377) & 2 (.656) \\
\hline
\end{tabular}} 
\end{table}

\bigskip

\section{Discussion and Conclusions}

Results of the analysis of two datasets consisting of individuals born prematurely as well as healthy controls suggest that in terms of gray matter variance the separation between males and females becomes increasingly larger as the brain transitions from adolescence to adulthood. 

The specific brain regions associated to variance effects at $19$ years are predominantly involved in cognitive processes including language (i.e. supramarginal gyrus, middle frontal gyrus), high-order executive functions (frontal pole, anterior cingulate gyrus) and emotion processing (superior temporal sulcus), with the exception of one limbic region (nucleus accumbens). Results further show a lateral asymmetry in sex differences, which merit further investigation. However, at age $30$ years the top ranking brain regions associated to variance components in both datasets mainly included limbic areas, amygdala, hippocampus, and nucleus accumbens which modulate affective processing, the pericalcarine cortex (or primary visual cortex V1), the pars triangularis (or Brodmann area $45$ which comprises Broca's area), which is central to semantic processing and the temporo-parietal junction, which has been associated with switching between self and other representations \cite{sowden15}. 

We noticed however some differences in the specific brain regions that separate males and females between preterm individuals and controls at both 19 and 30 years. While the top two ranking regions at 19 years are the same for both groups, the preterm cohort displays greater sex separation first in executive areas (frontal pole and anterior cingulate gyrus) and then in areas implicated in theory of mind (temporo-parietal junction) \cite{apperly04}, and the ranking is reversed in controls. At age 30 the ranking becomes even more different between the groups. In the preterm cohort theory of mind areas ranks higher than areas belonging to the mesolimbic pathway which sub-serves reward and motivation (nucleus accumbens, hippocampus and amygdala), while in controls the order of importance is reversed and areas belonging to the mesolimbic pathway rank higher than theory of mind regions. Other areas ranking higher in controls compared to preterm adults are language (pars triangularis) and visual regions (pericalcarine sulcus). These findings are consistent with the results of previous studies which reported sex-specific structural differences in preterm adolescents relative to controls \cite{constable08, scott11} and contribute to an increased understanding of specific psychiatric vulnerabilities in the different groups. For example, it is acknowledged that preterm adults are less likely to engage in risk taking behaviours than controls \cite{hack06}, but are more likely to exhibit a ``global withdrawn personality'' \cite{madzwamuse15}, which shares some of the characteristics of autism including impairments in social competence and theory of mind.

In terms of regional specificity, the areas that display the greatest contribution in explaining age-specific variability between males and females in our analysis show some consistency with regions showing significant sex bias in voxel-based mean estimates of brain volume and density as reported in Ruigrok's meta-analysis \cite{ruigrok14}. The results of our study have potential implications for the understanding of differential expression of several neuropsychiatric conditions and their cognitive manifestation in men and women, as well as their neuroanatomical underpinnings. As the involvement of limbic brain areas in explaining age-specific variability is most evident in the $30$ year old group, our results could be particularly important for the elucidation of sex-biased ages of onset for neuropsychiatric conditions, which display structural and functional alterations to the limbic system, such as depression \cite{wessa15} and schizophrenia \cite{okada16}.  

\bigskip

\renewcommand*{\bibfont}{\footnotesize}
\bibliographystyle{IEEEtran}
\bibliography{prni2016_ref}

\end{document}